\documentclass[preprint,sort&compress,5p]{elsarticle}

\usepackage{graphicx}
\usepackage[T1]{fontenc}
\usepackage{amssymb}
\usepackage{amsmath}
\usepackage{color}
\usepackage[utf8]{inputenc}

\newcommand{\bn}[1]{\mbox{\boldmath $#1$}}

\newcommand{\mb}{\mbox}

\journal{Physica E}

\begin{document}

\begin{frontmatter}

\title{Giant conductance and phase time anomalous events of hole quantum
transport.}

\author[label1]{S. Arias-Laso}
\ead{sariaslaso@gmail.com}
\author[label2]{L. Diago-Cisneros}
\ead{ldiago@fisica.uh.cu}

\address[label1]{Departamento de Física, ISPJAE, C.P. 19390, La Habana, Cuba.}
\address[label2]{Departamento de Física Aplicada, Facultad de Física, Universidad de La
Habana, C.P.$10400$, Cuba.}

\begin{abstract}
Events of giant conductance and anomalies of the phase transmission time for holes, are theoretically investigated within the multicomponent scattering approach. Based on this model, new analytical expressions for unitarity relations in the uncoupled hole transport are obtained and directly applied to study the behavior of the conductance and the phase transmission time in a double barrier resonant tunneling (DBRT) and a superlattice
$GaAs$-cladding layer$/(AlAs/GaAs)^{n}/GaAs$-cladding layer. Clear-signature evidences of giant conductance phenomena for hole transmission without valence-band mixing through a DBRT and a superlattice were found. The giant conductance effect losses robustness by manipulating the number of superlattice layers and by including the valence-band particles coupling as well. Phase time through the heterostructure exhibits extremal dependencies in the gaps and in the barriers, as those reported before for electrons. We have detected an earlier arrival phase time for the propagation of both flavors of holes within the barrier, in the order of few tenths of picoseconds. An appealing filter-like effect is presented, whenever a selective confinement strength arises independently for both flavors of holes in the uncoupled regime. Our results also prescribe noticeable evidences for both uncoupled and coupled hole fluxes, similar to those foretold by Hartman, upon transmission of electrons through opaque barriers.
\end{abstract}
\date{\today}

\end{frontmatter}

\section{Introduction}
\label{intro}

The impressive development of low-dimensional electronic and optoelectronic devices,
brought a new urgency to the essential measurement and modeling of charge carriers
transmission time and particularly of the tunneling time through specific potential
regions. When electrons and holes are involved, the low-dimensional device response
depends on the slower charge-carrier's traveling time \cite{Schneider89}. Undoubtedly, for
such systems it is crucial to study the tunneling time of holes. An equally important
reason for address investigations on heavy holes $(hh)$ and light holes $(lh)$, is their
intrinsic band mixing effects. Since the hole's transport is essentially an
interconnected-multichannel process, the theoretical calculation of transmission
quantities is in some sense more cumbersome than that for electrons.

Currently, the research of under-synchronic ano\-ma\-lous scattering events, is one of the major areas in the quantum transport
theory\cite{Drag03,PRB74,RMexPhaseTime,PPPEMTheory,WinHartmanEffect,WinNoSuperluminal}.
Regarding the long-standing controversy around superluminal velocities and total
transmission inconsistency, together with their wide-known ano\-malous character
\cite{Hauge,Landauer94}, they are welcome due to all possible benefits they open up for
design and development of nano-optoelectronic
devices\cite{Steinberg93,Spielmann94,Drag03}. There is a large accumulated knowledge about these topics -mostly dedicated to electrons, optical pulses and electromagnetic waves
\cite{PPP09,PRLsuperlatticetimes,Steinberg93,Spielmann94,PRL48,PRA28,PPPEMTheory}-,
accurately enough to unambiguously provide solid grounds for superluminal propagation, the paradoxical Hartman premonition and legitimacy of the phase time approach. Worthwhile noticing however that, hole transport analogous phenomena has been less studied yet.

Resonant mechanisms between propagating mo\-des oscillations and the self oscillations of the system are the fundamental cause, which gives rise to an isolated maximums of
transmission, often referred to as giant conductance (GG) phenomena. Events of GG have
been observed for electrons and optical pulses in many finite periodic systems, which
possess time reversal symmetry, namely: ballistic mesoscopic conductors, luminal pulses
and superlattices \cite{GGKadig1,GGPereyra,GGAllsopp,Romo1,Romo2}. Recent theoretical
studies in molecular devices suggested that could rise up particularly large values for
dynamic conductance \cite{GGMolecDevices} and novel experimental achievements pointed out
to a giant enhancement of electronic tunneling in a two-dimensional lattice of
graphene\cite{giantGgraphene, Geim07}. Lately, was found a strong dependence of the
conductance fluctuations with the density of states in graphene nano\-rib\-bons, which
represent an alternative insight to study the band structure of these physical
systems\cite{GNR}.

Owing to its relevance for quantum transport, the tunneling time has played a key role for a correct evaluation of electronic devices \cite{Hauge}. A large number of proposals in reply the challenge to settle how long does it takes to a particle to tunnel a single potential barrier \cite{MacColl}, have been put forward \cite{Hauge,Landauer94,But-Lan82}.
The lively and long standing debate concerning this matter
\cite{MacColl,BohmQT,WignerTime,Smith60,Baz67,But-Lan82,Buttiker83,Hauge,Landauer94}
-presumed not to finish completely-, it seems to have finally found its cornerstone in
precise experimental measurements for photons and optical pulses
\cite{Steinberg93,Spielmann94,Nimtz02,Ranfagni91,Ender92}. These genuine measurements put to the test, before all else, the very question loosely formulated by McColl and directly
bear on truly temporal features of tunneling phenomenon. Besides, they are consistent with
the phase time conception derived from group delay within the stationary-phase method
\cite{Hartman,BohmQT,WignerTime}. An additional motivation was the accurate description of
above mentioned experiments by careful phase-time calculations
\cite{PRLsuperlatticetimes,PPP09}.

In this paper we will present numerical calculations of the conductance and the phase time for an arbitrary number of alternate layers of $III-V$ materials, in the framework of the multicomponent scattering approach (MSA)\cite{PRB74}. The main idea is to search evidences of anomalous events in $hh$ and $lh$ transmission. In Sec.\ref{MSA} some theoretical remarks  within the MSA are exposed. Brief complementary insights to the uncoupled regime formula of scattering matrix unitarity relations and phase time will also be presented. Analytical expressions for the phase time and the conductance are applied in Sec.\ref{NumericalResults} to study the anomalous events associated with these physical quantities in the double barrier resonant tunneling (DBRT) and the superlattice. Phase time behavior for holes shows a very good agreement with the previous results for electron-tunneling through superlattices\cite{PRLsuperlatticetimes}. Evidences of the
giant conductance phenomena for uncoupled hole transport are shown. Finally, in
Sec.\ref{conclusions} we summarize and conclude.

\section{Conduction and phase time in the multicomponent scattering approach}
\label{MSA}

The MSA is based on standard $(N \times N)$ \mbox{\boldmath $k\cdot p\;$} effective
Hamiltonians and the multichannel transfer matrix (TM) method. In this approach the scattering amplitudes can be obtained straightforwardly \cite{PRB74}. The main purpose here, is to evaluate quantities of synchronic multi-mode transport for massive charge carriers moving through multi-layered structures at coupled and decoupled channel regimes.

Owing to consistency, we found useful to recall few basic outlines of the Kohn-L\"uttinger (KL) two-band model \cite{KL1955}, due to its widely accepted accuracy for describing dynamics of elementary excitations as well as electronic properties in the valence band, within the framework of the multiband effective mass approach. This model, is highly common regarding to the field of hole quantum transport through periodic heterostructures. It considers in the $\bn{k}\cdot\bn{p}$ approximation the highest two valence bands degenerated in the $\Gamma$-point of the Brillouin Zone. The usual $(4 \times 4)$ KL Hamiltonian \cite{KL1955} in the Broido-Sham representation \cite{PRB74}, has the form
\begin{equation}
\label{4x4Hkl}
 \begin{array}{ccc}
  \hat{\bn{H}}_{\textsc {kl}} & = &
  \left(
   \begin{array}{cccc}
     H_{11} & H_{12} & H_{13} & 0 \\
     H_{12}^{*} & H_{22} & 0 & -H_{13} \\
     H_{13}^{*} & 0 & H_{22} & H_{12} \\
     0 & -H_{13}^{*} & H_{12}^{*} & H_{11} \\
   \end{array}
  \right),
 \end{array}
\end{equation}
\noindent with
\begin{eqnarray*}
  H_{11} & = & A_{1}\kappa_{\textsc t}^{2} + V(z) - B_{2}\frac{\partial^{2}}{\partial z^{2}}, \\
  H_{12} & = & \frac{\hbar^{2}\sqrt{3}}{2\,m_{0}}\left(\gamma_{2}(k_{y}^{2}-k_{x}^{2}) + 2i\gamma_{3}k_{x}k_{y}\right), \\
  H_{13} & = & i\frac{\hbar^{2}\sqrt{3}}{2\,m_{0}}\gamma_{3}(k_{x}-ik_{y})\frac{\partial}{\partial z}, \\
  H_{22} & = & A_{2}\kappa_{\textsc t}^{2} + V(z) - B_{1}\frac{\partial^{2}}{\partial z^{2}},
\end{eqnarray*}
\noindent being
\begin{eqnarray*}
  A_{1,2} &  = & (\gamma_{1} \pm \gamma_{2}) \\
  B_{1,2} & = & (\gamma_{1} \pm 2\gamma_{2}),
\end{eqnarray*}
\noindent in atomic units. The matrix elements $H_{ij}$ show a direct dependency of the semi-empirical parameters of L\"{u}ttinger $\gamma_{i}\ (i=1,2,3)$, that typify each slab of the heterostructure. While $V(z)$ and $\kappa_{\textsc t}$, stand for the periodic potential, and the two-components in-plane \textit{quasi}-momentum, respectively \cite{PRB74}.

\begin{figure}[t]
  \centering
  \includegraphics[width=8cm]{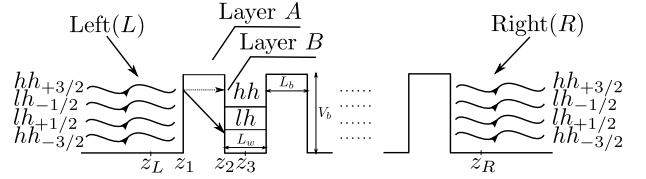}
   \caption{\label{fig:heterostructure}Schematic illustration of multichannel
$hh$ and $lh$ quantum transport processes through a $GaAs-\mathrm{cladding\
layer(L)}/(AlAs/GaAs)^{n}/GaAs-\mathrm{cladding\ layer(R)}$ superlattice.}
\end{figure}

To study the quantum transport properties of holes through semiconductor heterostructure,
we consider the system shown in Fig.(\ref{fig:heterostructure}), described by the
Kohn-L\"{u}ttinger Hamiltonian in the expression (\ref{4x4Hkl}). The four
accessible channels of the system, denoted as: $channel\ 1:\ hh_{+3/2},\\ channel\ 2:\
lh_{-1/2},\ channel\ 3:\ lh_{+1/2}$, and $channel\ 4:\ hh_{-3/2}$, are considered
simultaneously, following one of the bare bones of the MSA. We emphasize this upmost
advantage, as it allows clearly identify inter- and intra-subband hole transitions
together with transitions from Kramer-up toward Kramer-down states to time reversal, and
the other way around. The latter is not the usual case. In this approach, the segment
between $z_{L}$ and $z_{3}$ in Fig. (\ref{fig:heterostructure}), is defined as a
\emph{cell} and represents the period of the heterostructure superlattice \cite{PRB74}.

Combining the $(4\ \times\ 4)$ Kohn-L\"{u}ttinger Hamiltonian with the TM formalism, a
direct connection with the scattering theory is established \cite{PRB74}, then we have
$\bn{M}_{sv}(z_{R},z_{L}) = {\cal N}^{-1}\bn{M}_{fd}(z_{R},z_{L}){\cal N}$. Here,
$\cal{N}$ defines a relation between the traveling wave vectors of the tunneling problem
and their derivatives, with the $N-$dimensional state vectors that describe a propagating
or evanescent mode. This crucial transformation of the MSA, connects two different kinds
of TM. The first class of TM, $\bn{M}_{fd}(z_{R},z_{L})$, relates the linearly independent state functions and their derivatives ($fd$) between two points of the heterostructure in the form $\bn{M}_{fd}(z_{R},z_{L}) = \bn{N}(z_{R})\cdot\bn{N}(z_{L})^{-1}$, where the matrix
$\bn{N}(z_{i})$ can be constructed once the linearly independent solutions are
obtained~\cite{PhDLeo}. The second kind of TM, $\bn{M}_{sv}(z_{R},z_{L})$, whose elements
describe a propagating or evanescent mode depending on the energy, connects the state
vectors ($sv$) in two different points of the system. The mathematical structure of these
two matrices is conditioned by requirements of time-reversal invariance and flux
conservation \cite{MelloPereyraKumar}. A very strong criterium to verify the calculations
is based in the determinant of these transfer matrices which must be constant and equal to the unity, mostly.

Then applying the well known relations between the $\bn{M_{sv}} =
\left(\begin{array}{cc}
 \bn{\alpha} & \bn{\beta} \\ \bn{\gamma} & \bn{\delta}
\end{array}\right)$, and the scattering matrix \\ $\bn{S} =
\left(\begin{array}{cc}
 \bn{r} & \bn{t'} \\ \bn{t} & \bn{r'}
\end{array}\right)$ blocks \cite{MelloPereyraKumar}, one can directly evaluate the
relevant physical quantities of the scattered transport theory. Initially we have to
obtain the complex transmission amplitudes \cite{PRB74}
\begin{equation}
\label{transmission_amplitudes}
\begin{array}{c}
\bn{t} = \bn{\alpha} - \bn{\beta}\bn{\delta}^{-1}\bn{\gamma}, \\
t_{ij}=t_{R_{ij}}+ \;it_{I_{ij}},
\end{array}
\end{equation}
\noindent for each transition from the $j$-\emph{th} incoming channel (incident
propagating mode) to the $i$-\emph{th} outgoing channel (transmitted or reflected
propagating modes). It is worth noticing that within the MSA, none of the incoming
amplitudes is required to cancel, as it is commonly done. Hence, an authentic multichannel
and synchronic description of the hole transmission processes is naturally achieved. This
is to say that, we are able to distinguish accurately the contribution from each input
channel to the scattering process, modulated by the simultaneous presence of the rest of
the channels, which are, in principle, equally accessible to the incident hole stream.
None of the usual simplifications of the original \mbox{\boldmath $k\cdot p\,$}
Hamiltonians are invoked, thus the time reversal symmetry is assured. Given the
transmission amplitudes $t_{ij}$, it is possible to evaluate the total transmission
probability to channel $i$
\begin{equation}
 G_{i} = (e^2/\pi \hbar)\,\sum_{j} \left\vert t_{ij}\right\vert^{2},
 \label{Gi}
\end{equation}
the total two-probe Landauer conductance defined as~\cite{PRB74,Imry99}
\begin{equation}
 G =  (e^2/\pi \hbar)\,\sum_{i,j} \left\vert t_{ij}\right\vert^{2},
 \label{GT}
\end{equation}
and the transmission amplitude phase for each of the $16$ available paths
$ij$ \cite{PRB74}, $\theta_{ij} = \arctan\{t_{I_{ij}}/t_{R_{ij}}\}$. The
transmission phase time is obtained from \cite{PRB74}
\begin{equation}
\tau_{ij} = \hbar\frac{\partial}{\partial E}\theta_{nij},
 \label{phase_time}
\end{equation}
\noindent whose advantages and applicability criteria has been widely discussed in the
literature \cite{PRB74,PPP09,PRLsuperlatticetimes,Spielmann94,Heberle}. The sums in
equations (\ref{Gi}) and (\ref{GT}) run over all the open channels and the expressions
(\ref{Gi})-(\ref{phase_time}) and (\ref{phase_time_prop}) are defined for the $n$ cells
periodic structure showed in Fig.(\ref{fig:heterostructure}). Hereafter $E$ stands for the incoming particle's energy.

Tunable parameter $\kappa_{\textsc t}$, is essential to modify the in-plane dynamics of holes, \textit{i.e.} the degree of freedom transverse to the main transport direction, also refereed as the valence-band mixing for holes. This later is crucially relevant in order to understand the main problem envisioned here, and how we face it within the MSA. Whenever the \textit{quasi}-momentum parallel to the interfaces, $\kappa_{\textsc t} \approx 0$, the crossed transitions are not allowed and the system is said to yield an uncoupled hole regime. Under this particular condition, only direct incoming-outgoing channels are connected, therefore $(4\times 4)$ blocks $\bn{\alpha}$, $\bn{\beta}$, $\bn{\gamma}$ and $\bn{\delta}$ of the TM $\bn{M}_{sv}$ are diagonal matrices. The later, straightforwardly leads us to new  unitarity relationships, that satisfy the blocks of the multiband scattering matrix \bn{S} under an uncoupled hole regime, acquiring a more simple form:
\begin{equation}
\begin{array}{ccc}
\bn{\alpha}^{*}\bn{\alpha} - \bn{\beta}^{*}\bn{\beta} & = & \bn{I}_{N'} \\
\bn{\beta}^{*}\bn{\delta} - \bn{\gamma}\bn{\alpha}^{*} & = & \bn{O}_{N'} \\
\bn{\beta}\bn{\delta}^{*} - \bn{\gamma}^{*}\bn{\alpha} & = & \bn{O}_{N'} \\
\bn{\delta}\bn{\delta}^{*} - \bn{\gamma}\bn{\gamma}^{*} & = & \bn{I}_{N'},
\end{array}
\label{unit_desacoplados}
\end{equation}
where $N'=N/2$. Consequently, transmission and reflection amplitudes have reduced
expressions respect to (\ref{transmission_amplitudes}):
\begin{equation}
\begin{array}{lll}
 \bn{t} = 1/\bn{\alpha}^{*} & \mbox{with} & \alpha_{ii} = \alpha_{R_{ii}} +
\imath\alpha_{I_{ii}} \\ \bn{r} = -\bn{\delta}^{-1}\bn{\gamma} & \mbox{with} &
\delta(\gamma)_{ii} = \delta(\gamma)_{R_{ii}} + \imath\delta(\gamma)_{I_{ii}}
\end{array}.
\label{transmission_reflection_amp}
\end{equation}
Rewriting the phase time (\ref{phase_time}) for this specific case, we obtain:
\begin{equation}
\tau_{ii} =
\frac{\hbar}{T_{ii}}(\alpha_{R_{ii}}\frac{\partial\alpha_{I_{ii}}}{\partial E}
- \alpha_{I_{ii}}\frac{\partial\alpha_{R_{ii}}}{\partial E}).
 \label{phase_time_prop}
\end{equation}
Expressions like (\ref{transmission_reflection_amp}) were previously reported for the
resonant tunneling and band mixing in multichannel superlattices of propagating electron
systems \cite{PRLsuperlatticetimes}.

\section{Numerical results and discussion: tunneling time and giant conductance}
\label{NumericalResults}

For the numerical calculations that will be presented here we shall consider our layered
media with barrier thickness $L_{b} = (10-180)$ \AA, and well width $L_{w} = (50,150)$
\AA. For the barrier height we will have either $V_{b}=0.498\ eV$ or $V_{b}=0.550\ eV$.
The cases where the in-plane wavenumber (parallel to the interfaces) $\kappa_{\textsc t}
\approx 0$, with vanishing band mixing effects, and the cases where $\kappa_{\textsc t}
\neq 0$, with coupling phenomena effects on the quantum
transport\cite{PRB74,KCR,Klimeck01} will be separately study. Focusing to study the
behavior of the phase time $\tau_{ii}$ and the free motion time $\tau_{f}= n
l_{c}m^{*}_{h}/\hbar (k_{z})_{h}$ (being $l_{c}$ the cell longitude, $m^{*}_{h}$ the
effective mass of holes and $(k_{z})_{h}$ the eigenvalues of the scattering problem solved
with the MSA\cite{PRB74}) as a function of the incoming energy $E_{i}$, the barriers
height and thickness were fixed at $V_{b} = 0.23\ eV$ and $L_{b} = 30$ \AA, respectively.
Both, a double barrier ($n = 2$ cells) and a superlattice of $n=8$ cells were considered.
For consistency with the reported results for electrons\cite{PRLsuperlatticetimes}, a
$30\%$ concentration of $Al$ was fixed inside the barrier: $(GaAs / Al_{0.3}Ga_{0.7}As /
GaAs)^{n}$.

\subsection{Uncoupled hole regime phenomena}
\label{uncoupled_tunneling}

\begin{figure}[t]
\includegraphics[width=7.5cm]{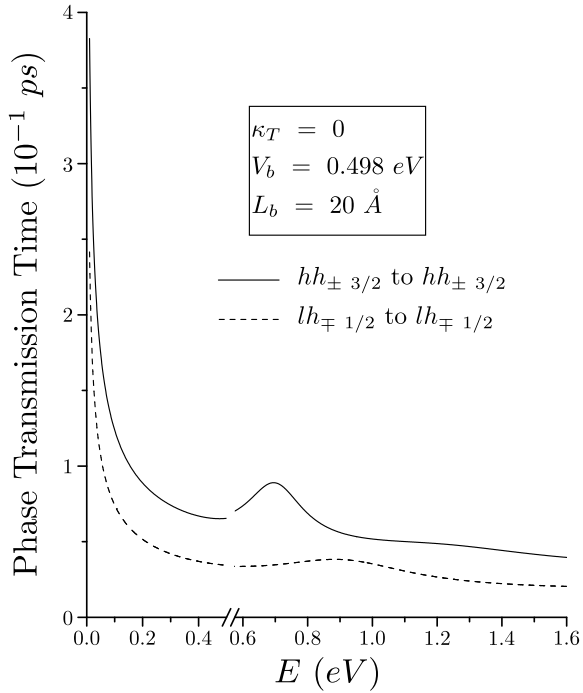}
\caption{\label{Fig2} Phase time $\tau_{ii}$ for a tunneling of holes through a single
barrier of $L_{b} = 20$ \AA $,$ as a function of the hole incident energy at
$\kappa_{\textsc t} = 0$ with $V_{b}= 0.498\ eV$. The solid (dashed) line stands for $i =
hh_{+3/2}\, (lh_{-1/2})$ direct path. In the broken interval a negative dip extents up to
$0.4$ ps.}
\end{figure}

In Figure \ref{Fig2}, we plot the phase time for $lh$ and $hh$ through a single potential
barrier. It is clear from this figure that the $lh$ modes (dashed line) are faster than
$hh$ modes (solid line), although the difference is not large. In an effort to validate
our numerical simulation with experimental data, we compare the results of Figure
\ref{Fig2} with those of Dragoman \textit{et. al.}~\cite{Drag03} The tunneling times
measured for holes, for null or small magnetic field in a ballistic-regimen nano-device,
are of the order of $10^{-13}\ s$. To reproduce that situation, we canceled the
interaction between holes by assuming $\kappa_{\textsc t} \approx 0$ (uncoupled regimen),
in the absence of magnetic field. In our calculation the phase time for $hh$ and $lh$
is also of the order of $10^{-13}\ s$, which is in an acceptable qualitative agreement
with experiment and it is just one order less than the traveling time for electrons
through a single barrier \cite{Landauer89}, as was detected experimentally for a DBRT
\cite{Heberle}. We have observed broaden peaks of the phase time for $hh_{+3/2}$ direct
transition at $ E \approx 0.7\ eV$ and for $lh_{-1/2}$ direct transition at $E \approx
0.9\ eV$. We have observed, although not plotted, that the transmission coefficient maxima
are located at comparable energies. It was shown in Ref.\cite{PRLsuperlatticetimes} that,
the tunneling time follows the band structure. This feature of hole tunneling agrees also
with calculations of Ref.\cite{NivelesWA}, where a particular case of the envelope
function approximation was used. Evaluating the tunneling times for energies in the range
$0.40\ eV \leqslant E \leqslant 0.60\ eV$ we have found, although not shown, a reduction
of $\tau_{ii}$ in $3$ orders of magnitude with respect to that shown in Figure \ref{Fig2}
when the in-plane momentum changes as $\Delta\kappa_{\textsc t} = 0.001$ \AA. This
behavior of $\tau_{ii}$, was earlier detected in experiments for a hole passage through a
slightly different system \cite{Ten96} -a double asymmetric quantum well (QW)-, and is
considered to arise from valence subbands mixing \cite{Ten96}.

\begin{figure}[t]
 \hspace*{-3mm}
 \includegraphics[width=7.5cm]{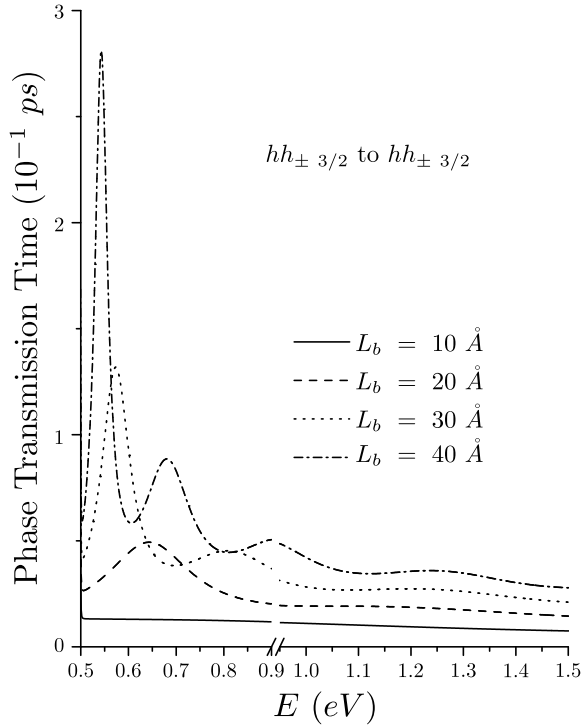}
\caption{\label{Fig3} Direct phase time $\tau_{ii}$ for hole tunneling through a single
barrier as a function of the hole incident energy for several values of the barrier
thickness at $\kappa_{\textsc {t}} \approx 0$, with $V_{b}= 0.498\ eV$ and $i = 1$.}
\end{figure}

Figure \ref{Fig3} displays a raising of the phase time while $L_{b}$ increases in the case
of uncoupled regime ($\kappa_{\textsc t} \approx 0$) of holes incidence. The same is
achieved for light holes, although not shown in this figure. This behavior agrees
qualitatively with the results of an experiment on time-resolved luminescence spectroscopy
\cite{Heberle}. For $E > 0.4\ eV$, $\tau_{ii}$ is sensitive to the raising of $L_{b}$ and
it is shown an oscillating behavior of the transmission phase time as function of the
incident energy. The resonances shift to smaller energies, increasing their intensities
with $L_{b}$. It is assumed understood that, when the de Broglie wavelength of the hole
propagating mode is a multiple of $L_{b}$, there is a resonance in the transmission and
correspondingly the phase time delay raises at these energies. As $L_{b}$ increases, new
resonances appear, thus the oscillating behavior is achieved. These oscillations
correspond to the quasi-stationary levels of the virtual QW with increasing width. In that
sense, from $\tau_{ii}$ properties we are able to forecast a longer lifetime for the lower
unbound hole levels. On general grounds, we found this in correspondence with the idea of
the quasi-stationary states lowering, whenever the virtual QW becomes wider.

\begin{figure}[t]
 \centering
 \includegraphics[width=7cm]{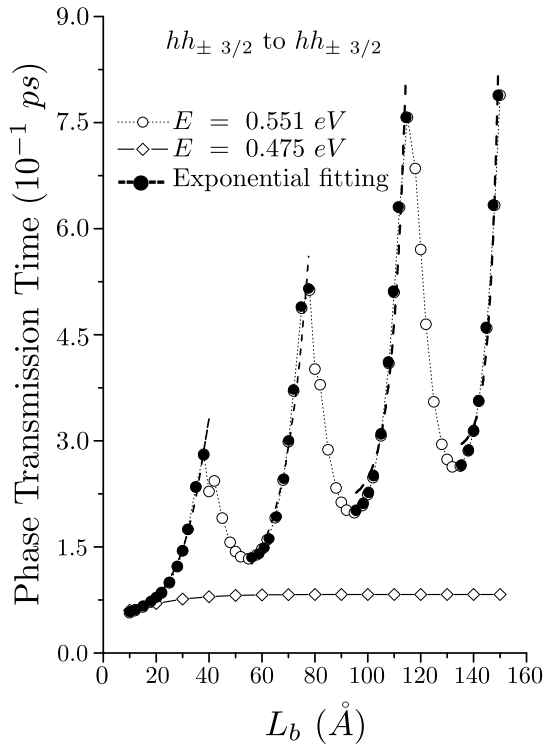}
\caption{\label{FigTauvsLbn1}Phase time \emph{versus} barrier thickness for the direct
transitions $hh_{\pm 3/2}\rightarrow hh_{\pm 3/2}$ given as solid diamonds/circles ($E <
V_{b}/E > V_{b}$), in the uncoupled case ($\kappa_{\textsc t}\approx 0$). A fragmented
exponential fitting is shown in dashed line (with full circles) at $E = 0.551\ eV$. The
barrier height $V_{b}=0.498\ eV$.}
\end{figure}

Figure \ref{FigTauvsLbn1} shows phase transmission time \emph{versus} barrier thickness
for the direct transitions $hh_{\pm 3/2}\rightarrow hh_{\pm 3/2}$ given as solid
diamonds/circles ($E < V_{b}/E > V_{b}$), in the uncoupled case. We prescribe appealing
tunneling events of holes for $E < V_{b}$ under uncoupled propagating regime. This figure
displays $\tau_{ii}$ for resonant direct $hh_{\pm 3/2}\rightarrow hh_{\pm 3/2}$
transitions (diamonds), which rapidly becomes autonomous from barrier broadening as
Hartman predicted for electrons \cite{Hartman}. These results are in accordance with the
transit time for optical pulses through opaque barriers (which transmit about $10^{-4}$ of
the incident radiation) \cite{Spielmann94} and also with theoretical predictions for the
delay time in one and two dimensions \cite{Steinberg94}. The circles in Figure
\ref{FigTauvsLbn1} correspond to numerical results within the MSA, for the tunneling time
of direct paths (indicated in the legend) for $E > V_{b}$. Here we must stress, that the
resonant oscillating behavior (on top of an average slope of smooth growth \cite{Com1}),
is an expected phenomena due to the barrier's interference effects with the continuum
states, wide accepted as Ramsauer-Townsend oscillations \cite{Penna06}. For $E > V_{b}$ we
briefly investigated segments in Figure \ref{FigTauvsLbn1}, pursuing a locally
exponential-like behavior of $\tau_{ii}$, as was suggested in a reported experiment with
holes \cite{Heberle}, within a rank comparable with their measurements. The interval of
barrier thickness length variance in their measurements, ranges from $30$ \AA $ $ up to
$80$ \AA. Notice in Figure \ref{FigTauvsLbn1} a sectional fitting, using an exponential
growth function \cite{Com2} (dashed line with full circles). We do believe that
oscillations are already intrinsically present in Heberle's, \textit{et al.} accurate
investigation \cite{Heberle} and could be shown by including more experimental points or
enlarging in rank the search.

\subsubsection*{Phase time through DBRT and superlattice}
\label{tauvsE}

\begin{figure}[t]
 \centering
 \includegraphics[width=7.5cm]{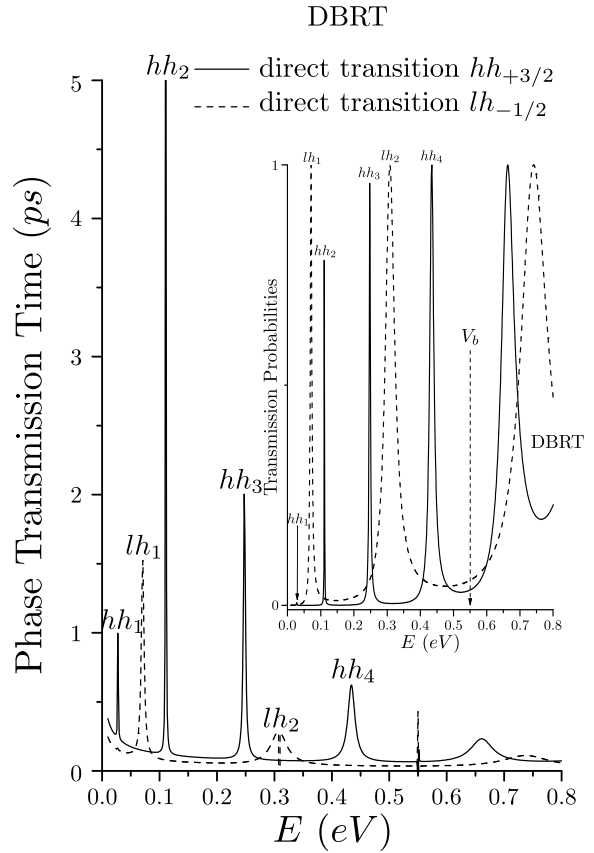}
 \caption{\label{Fig5}Direct phase time $\tau_{ii}$ for hole tunneling through a DBRT
structure as a function of the energy of an impinging hole stream at $\kappa_{\textsc t}
\approx 0$, with $V_b= 0.550$, $L_{b}=20$ \AA and $L_{w}=50$ \AA, for $i = hh_{+3/2}$
(solid line), and $i=lh_{-1/2}$ (dashed line). The inset displays transmission
coefficients for the same direct transitions, through an identical DBRT heterostructure,
as a function of the incident energy. The labelling for the hole levels follows that of
the Reference \cite{Klimeck01}.}
\end{figure}

Figure \ref{Fig5} is devoted to analyze briefly, the spectrum and the quasi-stationary
bound states lifetime in a DBRT. The curves of Figure \ref{Fig5} (inset), show several
transmission phase times $\tau_{ii}$ resonances for the uncoupled hole regime. In particular we have analyzed a semiconductor heterostructure of the form $GaAs/(AlAs/GaAs)^{2}/GaAs$. It
was shown in Ref. \cite{PRLsuperlatticetimes} that the tunneling time follows the band
structure profile. Figure \ref{Fig5} (inset) displays clearly, at $E < V_{b}$, the
familiar discrete quasi-stationary hole-level spectrum of the embedded QW, widely accepted
from hole quantum transport calculations
\cite{PRB74,Mori93,Erdogan93,Proetto95,NivelesWA,KCR,Rosseau89}. We have found at low
energies ($E < V_{b}$), that the $\tau_{ii}$ resonances are allocated at energies which
correctly reproduce the hole spectrum previously reported for the DBRT experiment
\cite{EWBL85}, as expected (see Table \ref{Tab1}). Notice the good quantitative agreement
of our approach in Table \ref{Tab1}, with similar values reported elsewhere \cite{EWBL85}.
A previous comparison \cite{PRB74} with a robust and widely accepted theoretical model
\cite{NivelesWA}, exhibits the advantages of the present approach.

\begin{table}
\caption{\label{Tab1} Comparison to typical quasi-steady DBRT hole-level
structure with $V_{b}=0,550\ eV$, $L_{b}=10$ \AA, and $L_{w}=50$
\AA.}
\begin{tabular}{lcc}
\\
\hline \hline
Levels \cite{Klimeck01}
& \multicolumn{2}{c}{Resonances $[eV]$} \\
& Experiment\cite{EWBL85}. & Theory (MSA). \\
\hline
$hh_{1}$ & 0,028 &  0,027 \\
$lh_{1}$ & 0,073 &  0,071 \\
$hh_{2}$ & 0,111 &  0,110 \\
$hh_{3}$ & 0,245 &  0,247 \\
$lh_{2}$ & 0,299 &  0,307 \\
$hh_{4}$ & 0,428 &  0,434 \\
\\
\hline \hline
\end{tabular}
\end{table}

Tunneling time can not be quantitatively determined from the resonant tunneling line
width,~\cite{Nimtz96} nevertheless some qualitative prognosis is allowed. Notice the
general trend of larger values for $hh$ states phase time in comparison to that of $lh$
states. This is in correspondence with the properties we have found for the
transmission-coefficient resonances depicted in the inset of Figure \ref{Fig5}, which
shows, as tendency, that $hh$ resonances are thinner than the $lh$ ones. This observation
foretells a stronger confinement for heavy holes, and we guess it could be used in
resonant-tunneling diode design and nano-modelling \cite{Klimeck01}.

Among the different formulations of tunneling time with recurrent experimental agreement
\cite{But-Lan82}, the phase time have clearly advantages. In Reference
\cite{PRLsuperlatticetimes}, was demonstrated that the phase time associated with the
passage (tunneling or not) of particles is a relevant quantity of the quantum transport,
which was calculated and successfully compared with precise experimental measurements
\cite{Steinberg93,Spielmann94}. Moreover in Reference \cite{PRLsuperlatticetimes}, some
appealing effects for electron phase time were predicted, namely: superluminal phase
propagation and resonant-band structure. In an attempt to distinguish similar behaviors
for holes, we perform some numerical simulations, exhibited in the next two figures.

\begin{figure}[t]
\centering
\includegraphics[width=8cm]{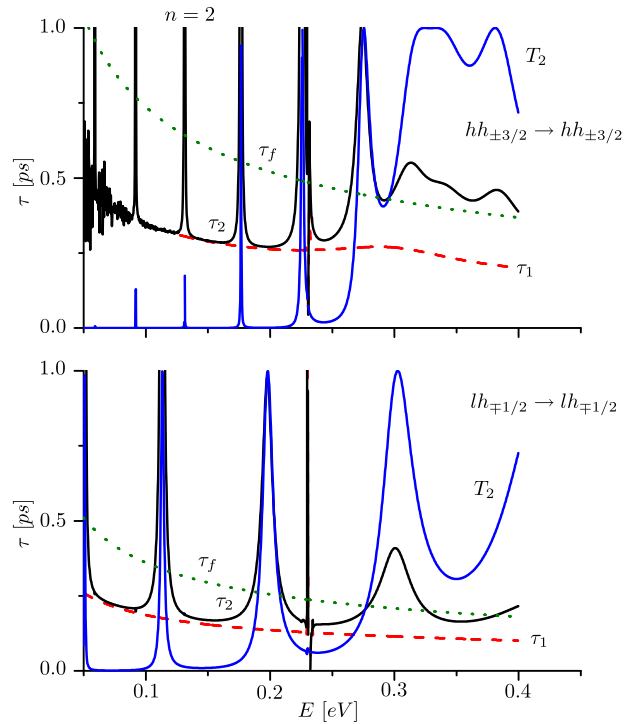}
 \caption{\label{FigTauvsEn2}Top(Bottom) panel shows the evolution of the phase
time, transmission probability, free motion time through $n=2$ cells (DBRT) and
the phase time through one single cell as function of the incoming energy for
direct transitions of $hh_{\pm\ 3/2}\left(lh_{\mp\ 1/2}\right)$. With $V_{b}=0.23\ eV$,
$L_{b}=30$ \AA and $L_{w}=150$ \AA. A $30\%$ concentration of $Al$ was fixed inside the
barrier.}
\end{figure}

In Figure \ref{FigTauvsEn2} the phase transmission time $\tau_{1(2)}$, through a single
cell ($n=1$) and a DBRT ($n=2$) respectively, together with $T_{2}$ are shown in
connection with the free motion time $\tau_{f}$ (dotted line). The last, is the
propagation time of the quasi-particle in absence of scattering potentials. The direct
paths under analysis, are indicated in the legends of each panel. A resonant behavior for
$\tau_{2}$ is very clear for both $hh$ states (top panel) and $lh$ states (bottom panel).
In the forbidden energy region, $\tau_{2}$ approaches $\tau_{1}$ (dashed red line), which
represents its lower bound as had been foretold for electrons \cite{PRLsuperlatticetimes}.
The relationship between $\tau_{2}$ and $T_{2}$ (blue line) resonances - as one can see
from (\ref{phase_time_prop})-, justify $\tau_{2}$ to accurately reproduce, the
quasi-stationary states of the embedded $GaAs$ QW. Another appealing feature arises for
forbidden energy regions ($T_{2}$ vanishes), where $\tau_{2} < \tau_{f}$. The opposite is
predicted for allowed energies. For $hh$ states (top panel) of incoming energies around
$100\; meV$, $(\tau_{f} - \tau_{2}) \approx 0.4\; ps$ , while for $lh$ states (bottom
panel) $(\tau_{f} - \tau_{2}) \approx 0.1\; ps$. This earlier arrival time for $hh$ and
$lh$, was predicted for electron tunneling through a same heterostructure
\cite{PRLsuperlatticetimes}, and suggest a more speedily passage of holes through a DBRT
like $(GaAs / Al_{0.3}Ga_{0.7}As / GaAs)^{2}$.

It remains to notice that $\left(\tau_{f}\right)_{hh}>\left(\tau_{f}\right)_{lh}$ for the
DBRT, having a bigger difference at low energies, with respect to that of high energies.
Another remarkable distinction is the width of $T_{2}$ resonances, which are noticeable
widest for $lh$. This difference yields to a larger confinement for $hh$ inside the
embedded QW, with respect to that for $lh$. Thus  greater lifetime for $hh$ in the
scattering region is expected.

\begin{figure}[t]
\centering
\includegraphics[width=8cm]{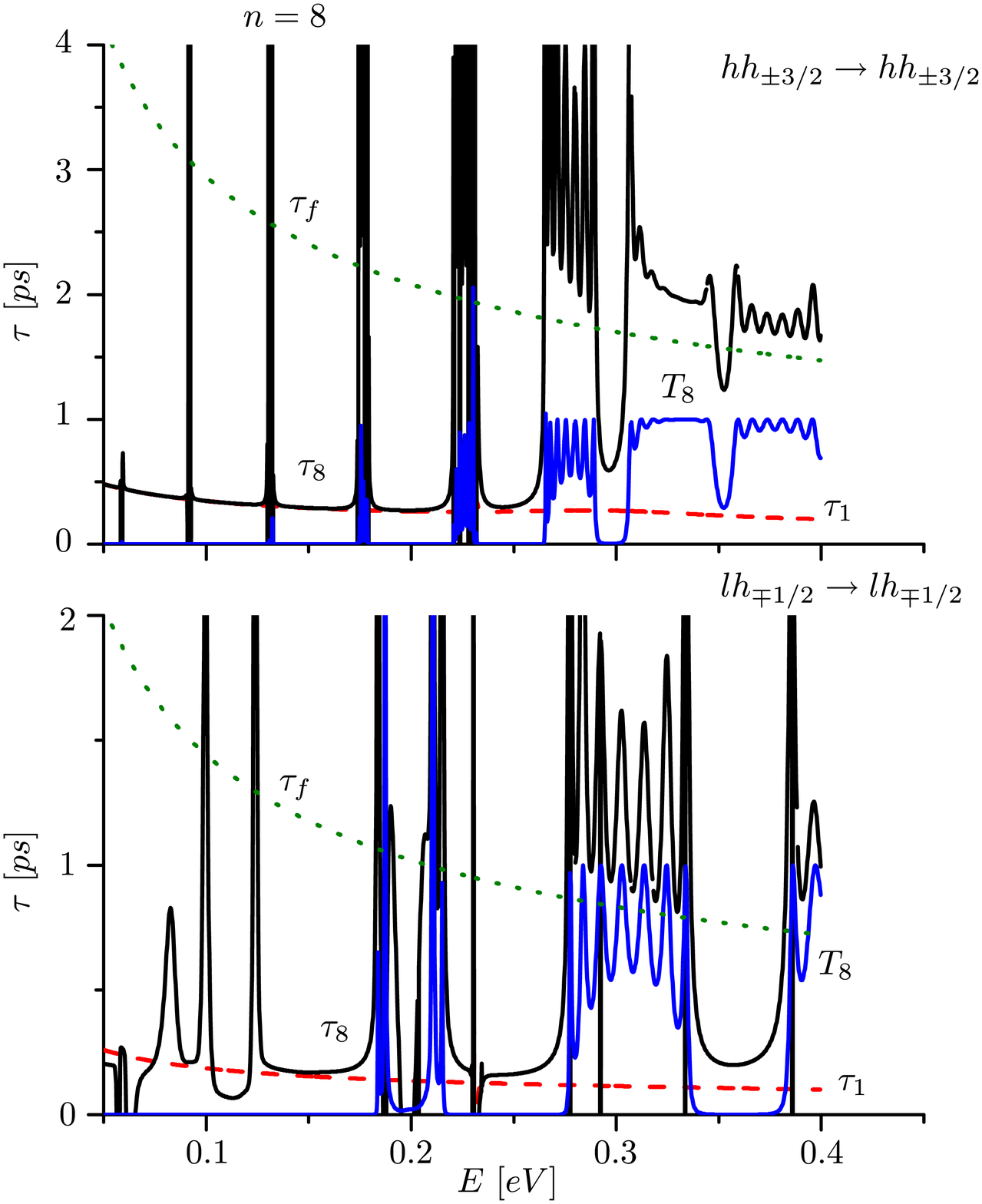}
 \caption{\label{FigTauvsEn8}Top(Bottom) panel shows the evolution of the phase
time, transmission probability, free motion time through $n=8$ cells (DBRT) and
the phase time through one single cell as function of the incoming energy for
direct transitions of $hh_{\pm\ 3/2}\left(lh_{\mp\ 1/2}\right)$. With $V_{b}=0.23\ eV$,
$L_{b}=30$ \AA and $L_{w}=150$ \AA. A $30\%$ concentration of $Al$ was fixed inside the
barrier.}
\end{figure}

In Figure \ref{FigTauvsEn8} the phase time $\tau_{8}$ and transmission coefficient
$T_{8}$, through a superlattice ($n=8$) undoubtedly follow a resonant-mini-band structure.
Direct hole transitions were considered and appear indicated in the legends of each panel.
For finite periodic systems -which is the case envisioned here-, a finite number of
intra-band energy levels \cite{PPP02} are typically observed, at variance  with the
continuous band structures predicted for infinite periodic systems. In the mini-gaps,
$T_{8}$ vanishes and $\tau_{8}$ tends to $\tau_{1}$, which is similar to the DBRT case
discussed before. On the other hand, in the mini-bands region $\tau_{8}$ exhibits a
resonant behavior approaching $\tau_{f}$, which behaves as a lower bound
\cite{PRLsuperlatticetimes}. In the scattering regions, $\tau_{8}$ increases as the
incoming energy approximates to any allowed energy level. While, in the forbidden regions
the holes propagation is away faster, reflecting the evanescent behavior and the lack of
hospitality of the barrier. These evidences for $hh$ and $lh$ phase time, resemble those
for electrons as was foretold for the tunneling time in a similar superlattice
\cite{PRLsuperlatticetimes}.

\begin{figure}[t]
\centering
\includegraphics[width=8cm]{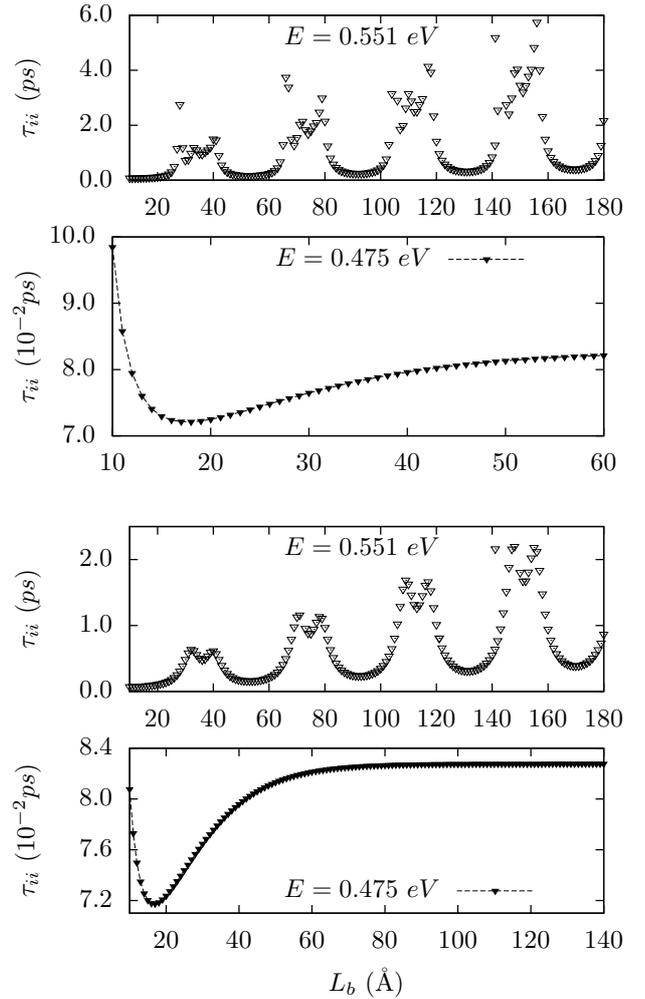}
 \caption{\label{FigTauvsLbn24} Phase transmission time as a function of the
barrier thickness for heavy hole transition $hh_{\pm 3/2} \rightarrow hh_{\pm
3/2}$ within the uncoupled regime, $\kappa_{T} \approx 0$. Both incident
energies $E$: less (full triangles) and greater (empty triangles) than the
barrier height $V_{b}=0.498\ eV$ were fixed at $0.475$ eV and $0.551$ eV, respectively.
The two bottom and top panels represents the numerical values of this time for the DBRT
and the superlattice of $n=4$ cells respectively.}
\end{figure}

Figure \ref{FigTauvsLbn24} exhibits two dissimilar behaviors of $\tau_{ii}$ for direct
transitions $hh_{\pm 3/2}\rightarrow hh_{\pm 3/2}$ as function of the barrier thickness.
For energies under the potential barrier (full triangles) $\tau_{ii}$ becomes independent
of the barrier thickness, being consistent with Hartman's classical premonition for
electrons \cite{Hartman}. In the DBRT, $\tau_{ii}$ becomes autonomous starting from
$L_{b}\approx 60$ \AA, meanwhile, in a superlattice of $n=4$ cells, this quantity
saturates at about $L_{b}\approx 50$ \AA. We can also note a decrease of the phase time in
the region $(10-15)$ \AA\ for the DBRT and around $(10-17)$ \AA\ for the superlattice,
showing the attractive effect of the barrier, as a reminiscence of the similar behavior
widely quoted in the literature for electrons through a simple barrier
\cite{Hauge,Landauer94}. For energies higher than the potential barrier (empty triangles )
$\tau_{ii}$, in the DBRT and in the superlattice, exhibits a resonant oscillating behavior
similar to those of the single barrier (see Figure \ref{FigTauvsLbn1}). This is a
consequence of an interference effect between the reflected wave and the continuum states,
known as Ramsauer-Townsend oscillations as commented above. By comparing Figure
\ref{FigTauvsLbn1} and Figure \ref{FigTauvsLbn24}, it is clear that oscillations
\cite{RMexPhaseTime,Penna06} remain robust for the DBRT. The robustness of this feature
vanishes for the superlattice, provides that well-resolved oscillating curves are no
longer observed for $n=4$ cells.

\subsubsection*{Giant conductance phenomena}
\label{GGphenomena}

\begin{figure}[ht]
\centering
\includegraphics[width=7cm]{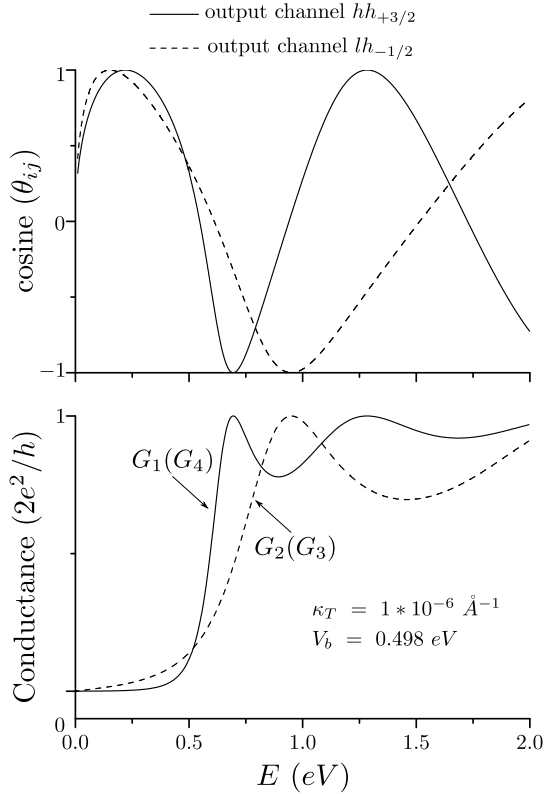}
 \caption{\label{Fig6} The bottom panel plots the $4$ outgoing channel
conductances $G_{j}$ through a single barrier heterostructure with $L_{b}=20$
\AA, where $i=1,4$ (solid line) represents the $hh_{\pm 3/2}$ channels,
respectively, and $i=2,3$ (dashed line) represents the $lh_{\pm 1/2}$ channels,
respectively. The top panel displays the cosine of phase shift for the same
transitions.}
\end{figure}

It is quite well established in the literature the strong relation between the
transmission magnitudes and the wave function's phase \cite{GGKadig1,GGPereyra,PPP02}.
Specifically, the giant conductance events are nearly related with phase shifts and phase interference phenomena \cite{GGPereyra}. It has been shown for various mesoscopic systems, the correlation between the variations of the conductance and an even
function of the phase shift, due to an alternatively appearance of maximum-minimum valued sharp peaks or dips, whenever an even functional of the phase shift, changes by almost $\pi$ radians \cite{GGKadig1,GGPereyra,Vegvar94,Pothier94,PetrashovPRL93}. Here, the cosine of the phase shift was chosen to verify, whether strike maximum-transmission peaks occur or not, in a negligible conductance background. Commonly, the $cos(\theta_{ij})$ is plotted, as it is a suitable even function to show evidences of giant conductance events \cite{GGKadig1,GGPereyra}.

Figure \ref{Fig6} (bottom panel), presents the conductance $G_j$ to direct channels
$hh_{\pm 3/2}\rightarrow hh_{\pm 3/2}$ (solid line), $lh_{\pm 1/2}\rightarrow lh_{\pm
1/2}$ (dashed line), and the cosine of phase shift for the same transitions (top panel)
for the uncoupled hole regime as function of the incident energy. It is clear
that the maximum values in the conductance for both output channels are always accompanied
by an extreme change in the cosine of the phase difference. One can see, that the curve
for $hh$ output channel (solid line) changes two times in $\pi$ radians and in the bottom
panel appear two $G_{j}$ maximum correspondingly (at $E\ \approx\ 0.72\ eV$ and at $E\
\approx\ 1.25\ eV$). Meanwhile for the $lh$ output channel (dashed line), there is only
one change (at $E \approx 0.98$ eV). Direct paths contributions are the largest ones, as
it follows from vanishing band mixing effects as $\kappa_{\textsc t}\approx 0$. The
maximum values shown in the bottom panel of Figure \ref{Fig6}, for an even function of
$\Delta\theta$ ($\cos\Delta\theta$), at even multiples of $\pi$
\cite{Vegvar94,Pothier94,GGKadig1,GGAllsopp,GGPereyra}, are not enough to confirm an
abrupt increment of transmission. Indeed, there is nothing particularly large about hole
transmission through a single cell, as that reported for electrons in finite periodic
systems with time reversal symmetry \cite{GGPereyra}. Besides, in the single-cell case,
does not exist a resonant mechanism that allows hole conductance to become giant through
the barrier.  However, it is tempting to have a fairly reference frame to compare to what
we will see next for a DBRT. Worthwhile noticing -although not shown here-, the largest
contribution of the imaginary scattering amplitude component, that give rise $G_{j}$ to
maximize in the same energy intervals, as we have shown in the Figure \ref{Fig6} bottom
panel.

\begin{figure}[ht]
\centering
\includegraphics[width=8cm]{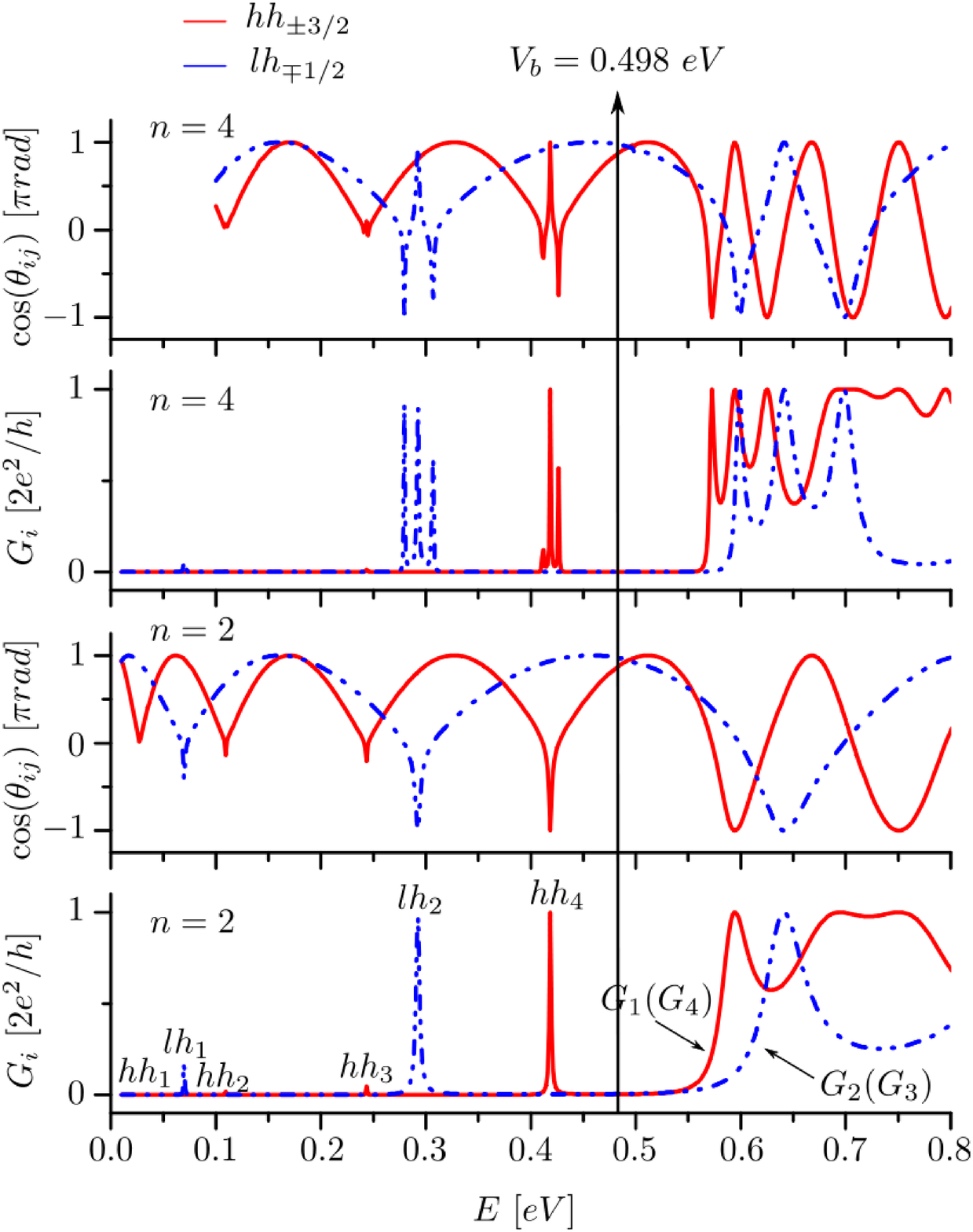}
 \caption{\label{FigGG} Dependency of the $4$ outgoing channels conductance and
the cosine of phase shift with hole incident energy, through a DBRT and a
superlattice with $n=4$ cells of width $L_{b}=20$ \AA$,$ and height
$V_{b}=0.498\ eV$.}
\end{figure}

Well-established time reversal symmetry of the $(4 \times 4)$ Kohn-L\"{u}ttinger model, is
not fulfilled by the $(2 \times 2)$ subspaces  as was recently demonstrated
\cite{LPRCScripta}. Events of GG have been observed in many finite periodic systems, which
possess time reversal invariance but as yet not observed in time asymmetric ones.
Probably, this explains why other accurate calculations made in the $(2 \times 2)$
Kohn-Luttinger subspaces, did not report GG events for holes
\cite{Krol95,Erdogan93,Heberle,Klimeck01}. Figure \ref{FigGG} displays the conductance and
phase shift behavior as a function of the incident energy, for direct transitions to
$hh_{\pm 3/2}$ (solid line) and $lh_{\pm 1/2}$ (dashed line) output channels. The $G_{i}$
curves exhibit evidences pointing out the existence of well-defined resonances for $E <
V_{b}$. It is relevant in this case, the presence of striking conductance events, as can
be clearly noticed in the panels for $n=2$ of Figure \ref{FigGG}. Indeed, peaks $lh_{2}$
and $hh_{4}$, have found to have a clear signature of GG phenomenon since: (i) they are
sharp resonances, (ii) they keep a simple relation with an even function
$cos(\theta_{ij})_{n=2}$ of the transmission phase $(\theta_{ij})_{n=2}$, thereby  peaks'
locations at energy, are coincident with those of $(\theta_{ij})_{n=2}$ whenever the last
changes by almost $\pi$ radians in a relative short energy interval
\cite{GGMolecDevices,GGKadig1,GGKadig2,GGPereyra,PPP02}, and finally (iii) there is a
strong phase interference mechanism between propagating or evanescent modes, giving rise
to resonant transmission \cite{GGPereyra}. As follows from the scattering theory, the GG
mechanism in a DBRT, derives from the interference between states, whenever the incoming
modes energy perfectly matches with one of the quasi-stationary states within the embedded
QW of the scatterer system. This agreement produces the essential conditions for energy to
maximize, \emph{i.e}. a resonant transmission occurs. Sharp peaks become visible at
energies $E \cong 0.29\ eV$ (dashed line) for direct transitions of $lh$, and for $E \cong
0.42\ eV$ (solid line) for direct transitions of $hh$. We have disregarded the $lh_{1}$
sharp resonances and others in the low energies region, because they do not reach maximum
and also the corresponding phase shifts differ significantly from $\pi$ radians. In
addition, it is necessary to mention the loss of the GG phenomenon in the superlattice. As
one can see in the top panels of Figure \ref{FigGG} (for $n=4$ and $E < V_{b}$), the
$(\theta_{ij})_{n=4}$ changes are not about $\pi$ radians. In this case the leak of
maximum transmission for the superlattice becomes visible. We conjecture the existence of
a competitive mechanism when the number of cells grow, and consequently a phase coherence
effect rises. For higher energies, the direct $G_{i}$ might give rise to a
\textit{plateau}, and we foretell the probable existence of another related striking
feature: the giant resistance. Despite a more ambiguous interference process in this case,
a sharp dips are expected as a consequence of a transmission rate decrease, due to a
resonant transfer of holes flux toward evanescent modes.

\begin{figure}[ht]
 \centering
\includegraphics[width=8cm]{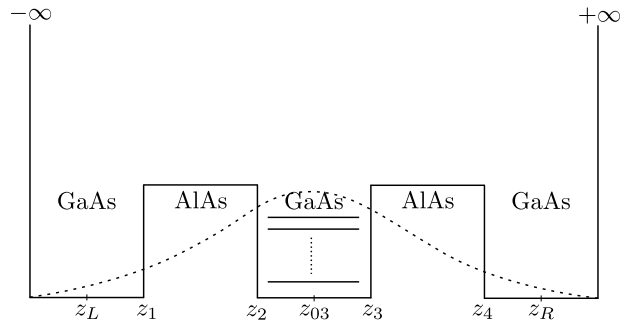}
 \caption{\label{FigDBRT} Schematic representation of the quasi-bound-state levels
(horizontal lines) and valence band lineup, within an infinite walls quantum well. The
ground state wave function (dotted line) is represented qualitatively. The energy is taken
as positive for convenience.}
\end{figure}

It is likely to underline that the energies of the resonances -whether they are giant or
not-, accurately reproduce the quasi-bounded hole spectrum as was published for the DBRT
\cite{NivelesWA}, were another approaching frame was applied. This is to say that, none GG
phenomenological analysis can possibly be complete if a resonant means is missing.  In
this sense, to endorse the GG evidences showed in Figure \ref{FigGG}, the most immediate
tool within the MSA, is the TM technique to estimate the hole states energies at the QW of
the DBRT. The basic idea here, is to search several states of the embedded finite quantum
well (fQW), by cautious tuning those of the external infinite quantum well (iQW). The last
we had achieved, by imposing that $z_{\textsc {l,r}}\,$ slowly tends to $z_{1,4}$,
respectively (see Figure \ref{FigDBRT}). This method approaches the iQw boundary
conditions to those of the fQW, and is a reasonably good sketch for an estimation of some
fQW levels.

Figure \ref{FigDBRT} shows schematically, holes quasi-bound-state levels (horizontal
lines) and valence band lineup, wi\-thin the iQW. For a numerical estimation of the
quasi-stationary levels within the fQW, we consider a cell enclosed between the $z_{L}$
and $z_{03}$ points. The barriers and well have the same dimensions of the Figure
\ref{FigGG}. Thus, it is undemanding and straightforward to reproduce a DBRT or a
superlattice, by a periodic concatenations of cells from $z_{03}$ \cite{PRB74}. The latter
considers facts of equivalence for points $z_{1}(z_{2})$ and $z_{3}(z_{4})$, respectively,
as they have the same materials to the left and to the right. As boundary condition for
the iQW, is standard to impose the wave function goes to zero at the points $(z_{\textsc
l}-\Delta z)$ and $(z_{\textsc r}+\Delta z)$
\begin{equation}
 \varphi(z_{\textsc l}-\Delta z)\equiv\varphi(z_{\textsc r}+\Delta z)\equiv\ 0,
\end{equation}
being $\Delta z$ the splitting distance between these points and the infinite potential.
In the framework of the TM approach, the latter requirement can be written as:
\begin{eqnarray}
 \nonumber
\left(\begin{array}{c}
  0 \\ \bn{\varphi}'(z)
\end{array}\right)_{(z_{\textsc r} + \Delta z)}  = \hspace{30mm} \\
\label{annulm_cond}
\hspace{-15mm}
\bn{M}_{fd}(z_{\textsc r} + \Delta z,z_{\textsc l} - \Delta z)\left(\begin{array}{c}
 0 \\ \bn{\varphi}'(z)
\end{array}\right)_{(z_{\textsc l} - \Delta z)}.
\end{eqnarray}
Therefore, the non trivial solutions of the expression (\ref{annulm_cond}) can be
numerically determined, by solving the equation
\begin{equation} \label{det}
  Det\{\bn{M}_{\textsc {AD}}(z_{\textsc r} + \Delta z,z_{\textsc l} - \Delta z)\}=0,
\end{equation}
\noindent being $\bn{M}_{\textsc  {ad}}$ the \emph{amplitude-derivative} matrix block of
the TM $\bn{M}_{fd}$. Up to this end, it is turn clear from Table \ref{Tab2}, a reasonably
good agreement between our estimation for the energies of several quasi-steady levels of
the fQW and the resonant-band structure displayed in Figure \ref{FigGG}. Now it is
undoubted, the very existence of these fQW levels, and therefore their role as a fair
resonant mechanism of the GG phenomena in the DBRT we claim for. The $lh_{2}$ and $hh_{4}$
energy values showed here, are a truthful prove of the GG phenomena, predicted for the
uncoupled hole transport in the DBRT for sharp resonances allocated at similar energies
(see Figure \ref{FigGG}).

\begin{figure}[t]
\vspace*{-1cm}
 \includegraphics[width=8cm]{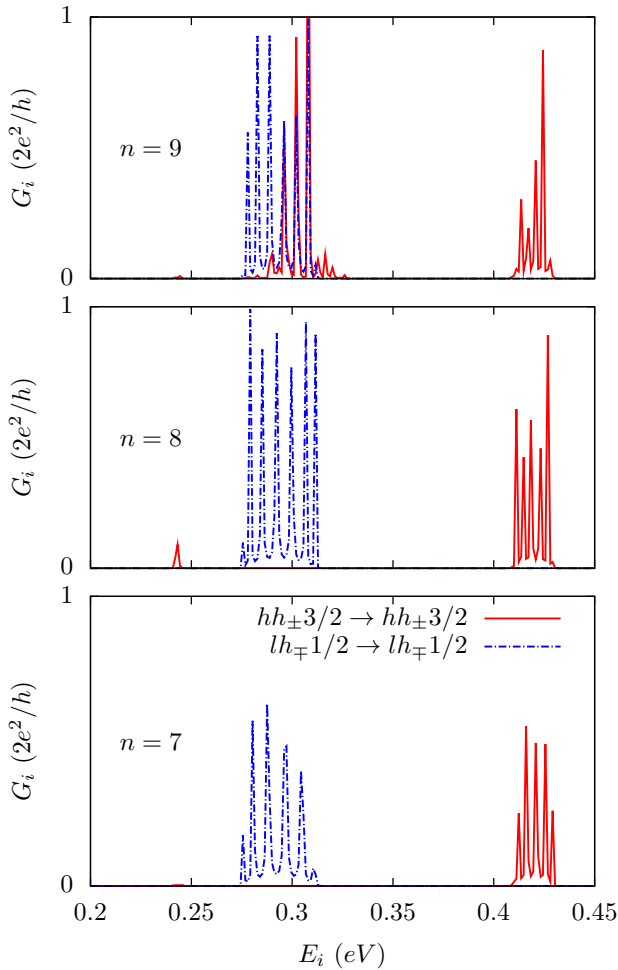}
\caption{\label{FigGn789} Two-probe conductance through the four direct accessible
channels as a function of the incoming energy on the $\kappa_{T} \approx 0$ limit, for a
superlattices of $n = 7,8,9$ cells (from bottom to top) in the range of energies between
$[0.2-0.45]\ eV$.}
\end{figure}

\begin{table}
 \caption{\label{Tab2} Quasi-steady fQW  hole-level for a DBRT structure, quoted by
Eq.(\ref{det}), with $V_{b}=0.498\ eV$, $L_{b}=20$ \AA, and $L_{w}=50$ \AA.}
\begin{tabular}{ll}
\\
\hline \hline
  Levels (Labeling from Ref.\cite{Klimeck01})
  & Resonances $[eV]$\\
  \hline
  $hh_{2}$ & 0.1362 \\
  $lh_{2}$ & 0,2978 \\
  $hh_{4}$ & 0,4112 \\
\\
\hline \hline
 \end{tabular}
\end{table}

\begin{figure}[ht]
\centering
\hspace*{-8mm}
 \includegraphics[width=8cm]{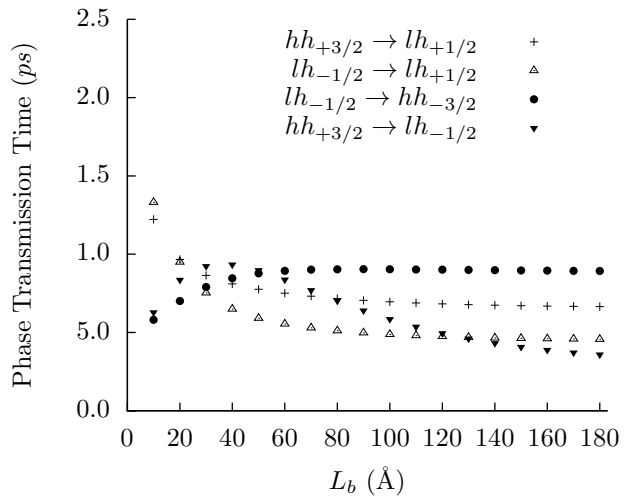}
\caption{\label{FigTauvsLb8akT510-4} Phase transmission time through one single cell
\emph{versus} barrier thickness for the direct and crossed passages of holes in a coupled
case ($\kappa_{T} = 5\times 10^{-4}$ \AA$^{-1}$) for $E=0.475\ eV < V_{b}=0.498\
eV$.}
\end{figure}

Figure \ref{FigGn789} presents the dependency of the conductance $G$ of $hh$ and $lh$
through a superlattices of $n = 7,8,9$ cells with the energy of the incident flux. The
holes travel along all four accessible direct channels $hh_{\pm 3/2}\rightarrow hh_{\pm
3/2}$ (solid line) and $lh_{\pm 1/2}\rightarrow lh_{\pm 1/2}$ (dotted-dashed line). It is
wide open visible the thinner peaks of the $hh$ resonances, respect to those of the $lh$
ones, since we are restricted on the uncoupled regime and thereby $hh$-spectrum states are
the closest due to their larger effective mass. Further outstanding behavior, is the
filter-like effect that can be observed on both flavors of holes in the selected region.
Indeed, the barriers become opaque for the $hh/lh$ states at low/high energies, leading to
a stronger confinement for the $hh/lh$ at low/high energies. This fact yields a
larger/shorter lifetime for $hh/lh$ inside the allowed region of the scatterer system for
low energy, at variance with the opposite for higher energies. Dealing with a superlattice
of $n=9$ cells (topmost panel), we found an $hh-lh$ overlap region at the vicinity of
$0.3$ eV. The phenomenology described in Figure \ref{FigGn789}, allows us hypothesize it
as concerns applications in hole-based electronics, pursuing maximum-minimum conductivity
response as function of the charge carrier effective mass.

\begin{figure}[ht]
 \centering
\includegraphics[width=8cm]{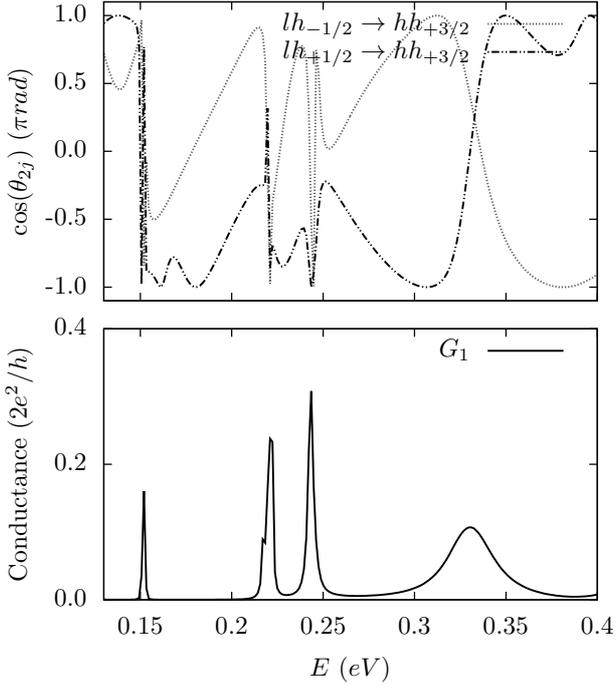}
 \caption{\label{FigCros1} The bottom panel shows the crossed-path contributions to the conductance through the first outgoing channel, $G_{1}$, for a DBRT with barriers of width $L_{b}=20$ \AA, and height $V_{b}=0.498\ eV$. The top panel displays the cosine of two phase shifts that match with the same outgoing channel, $hh_{+3/2}\rightarrow lh_{-1/2}$ and $hh_{+3/2}\rightarrow lh_{+1/2}$, both in the same range of energy.}
\end{figure}

\subsection{Coupled hole regime phenomena}
\label{coupled_tunneling}

\begin{figure}[ht]
 \centering
\includegraphics[width=8cm]{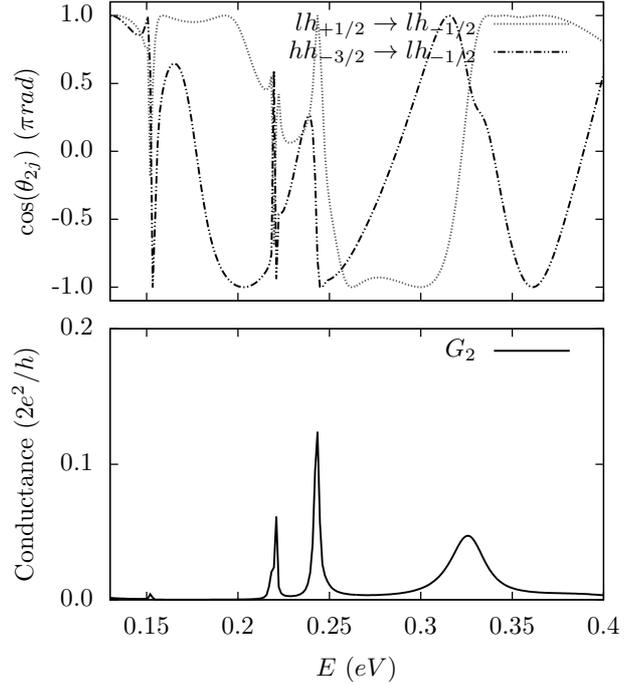}
 \caption{\label{FigCros2} The bottom panel shows the the crossed-path contributions to the conductance through the second outgoing channel, $G_{2}$, for a DBRT with barriers of width $L_{b}=20$ \AA, and height $V_{b}=0.498\ eV$. The top panel displays the cosine of two phase shifts that match with the same outgoing channel, $lh_{-1/2}\rightarrow lh_{+1/2}$ and $lh_{-1/2}\rightarrow hh_{-3/2}$, both in the same range of energy.}
\end{figure}

Let us considers first the Hartman effect under this regime. Figure \ref{FigTauvsLb8akT510-4} depicts the phase transmission time as a function of the barrier thickness for several crossed transitions paths, namely: $hh_{+3/2}\rightarrow
lh_{+1/2},\, lh_{-1/2}\rightarrow lh_{+1/2}, \\ lh_{-1/2}\rightarrow hh_{-3/2}\,$ and
$\,hh_{+3/2}\rightarrow lh_{-1/2}$, of the sixteen possible ones, through a single-cell
scatterer system. This plot shows the phase transmission time curves as a function of the
barrier thickness $L_{b}$, given fixed $E < V_{b}$. Notice the changes of $\tau_{ij}$
through each path while $L_{b}$ increases, until  $\tau_{ij}$ reaches saturation values.
This autonomous behavior of the phase time with $L_{b}$, unambiguously gives rise to the
paradoxical Hartman effect \cite{Hartman}, which remains robust, even for a relative
strong coupling of holes. Owing to simplicity we dropped the remain transitions, provided
they follow a similar tendency. Worthwhile underlying some characteristics of $\tau_{ij}$
transient interval for several paths. Crossed transitions $hh_{+3/2}\rightarrow lh_{+1/2}$
(+) and $lh_{-1/2}\rightarrow lh_{+1/2}$ (up empty triangles), show negative slope before
reach saturation. Meanwhile the crossed path $hh_{+3/2}\rightarrow lh_{-1/2}$ (down full
triangles), exhibits a double change of slope. These behaviors clearly departs (in the
transient interval) from the classical Hartman premonition for electrons and deserve more
investigation to shed light on this topic. In the presence of interference effects
($\kappa_{\textsc t}\neq 0$) at $E \approx 0.55\ eV$, we have found abrupt tunneling phase
changes when $\Delta\theta \approx \pi$ radians, which seems to be an anomalous behavior
of the phase time. In Figure \ref{FigTauvsLb8akT510-4}, we have
found an overlap of $\tau_{ij}$ for several
crossed transitions. For example: with $j:1(2) \rightarrow 3:i$ at $L_{b}\approx 20$ \AA, also with $j:1(2) \rightarrow 2(3):i$ at $L_{b}\approx [110,140]$ \AA, and with $j:1 \rightarrow 2(3):i$ at $L_{b}\approx [70,80]$ \AA. In some cases the effective mass changes, and the incoming channels belong to a so called Kramer-\textit{up} degenerated hole states \cite{PRB74}, while the outgoing channels are Kramer-\textit{low} ones. It can be readily observed, similar  $\tau_{ij}$ matches, for other crossed transitions at different values of $L_{b}$. These coincidences relate to the strong $hh-lh$ propagating modes mixing, to the simultaneous treatment of the $4$-coupled channels of the present approach, and to $hh-lh$ interaction with an effective potential $V_{eff}$ due the barrier broadening. Accordingly, whenever a match in $\tau_{ij}$ takes place, the time spent by the $hh$ and $lh$ in the classically forbidden region, shall be the same.

\begin{figure}[ht]
 \centering
\includegraphics[width=8cm]{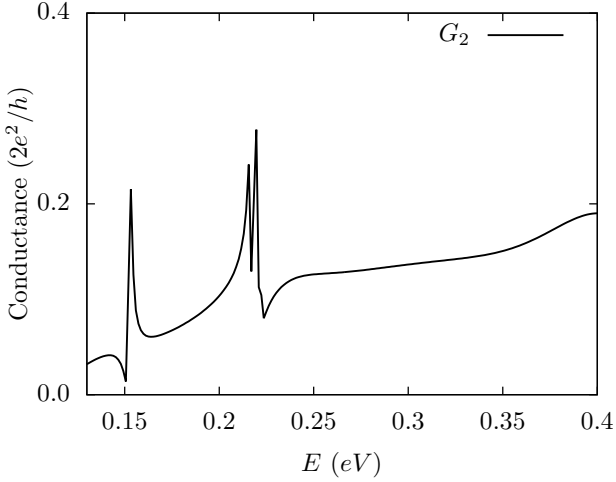}
 \caption{\label{FigCros4} Full contributions to the conductance through the second outgoing channel, $G_{2}$, for a DBRT with barriers of width $L_{b}=20$ \AA, and height $V_{b}=0.498\ eV$.}
\end{figure}

We turn now to discuss how robust are the hole giant conductance features, in the presence of valence-band mixing. Figure \ref{FigCros1} and Figure \ref{FigCros2} present the conductance and the $cos(\theta_{ij})_{n=2}$ through the outgoing channels $hh_{+3/2}$ and $lh_{-1/2}$, respectively, as a function of the incoming energy under interference between channels. We had taken $\kappa_{\textsc t}=10^{-3}$ \AA$^{-1}$. Despite the presence of sharp peaks at $E < V_{b}$ (qualitatively similar to those showed in Figure \ref{FigGG}), as one can observe at $E\cong 0.16\ eV$, $E\cong 0.22\ eV$ and $E\cong 0.24\ eV$ in Fig.\ref{FigCros1}, it is not seen a neat concordance with the widely accepted GG formulation posted above \cite{GGKadig1,GGPereyra,Vegvar94,Pothier94,PetrashovPRL93}, and illustrated in Fig.\ref{FigGG}. Firstly there is not a nice correspondence between the highest peak of $G_{1}$ at the vicinity of $E=0.24$ eV (see Fig.\ref{FigCros1} bottom panel) and the fluctuation of $(\theta_{ij})_{n=2}$, which for this peak is sensitive less than $\pi$ radians (see Fig.\ref{FigCros1} top panel). On the other hand, the values of $G_{1,2}$ exhibited in Fig.\ref{FigCros1} and Fig.\ref{FigCros2}, respectively, are in average almost one order smaller than those shown by Fig.\ref{FigGG}, being the comparative behavior observed in Fig.\ref{FigCros2} even more restrictive in this sense, with respect to Fig.\ref{FigGG}. Following these evidences, we conjecture so far, that the GG events are not straightforward signatures for hole tunneling traversing a DBRT, under a selected range of parameters in the coupled-particle regime.

Additional remarkable differences become clear from a comparison of this trend in the
coupled hole regime and Figure \ref{FigGG}. For instance, there is a small shift between
the position in energies of the peaks and the changes in the $cos(\theta_{ij})_{n=2}$. We guess that the last reflects the loss of maximum transmission when the valence-band mixing, \textit{i.e.} the interference between paths is incremented. Besides, the $G_{n}$ oscillations spectra in the low-energy interval shown in Fig.\ref{FigCros2}, seem to be of the Fano-type leading to assume the existence of interference of embedded-QW discrete levels with a continuum. The later gives rise to characteristically asymmetric peaks in transmission spectra \cite{Fano61}.

\begin{figure}[ht]
 \centering
\includegraphics[width=8cm]{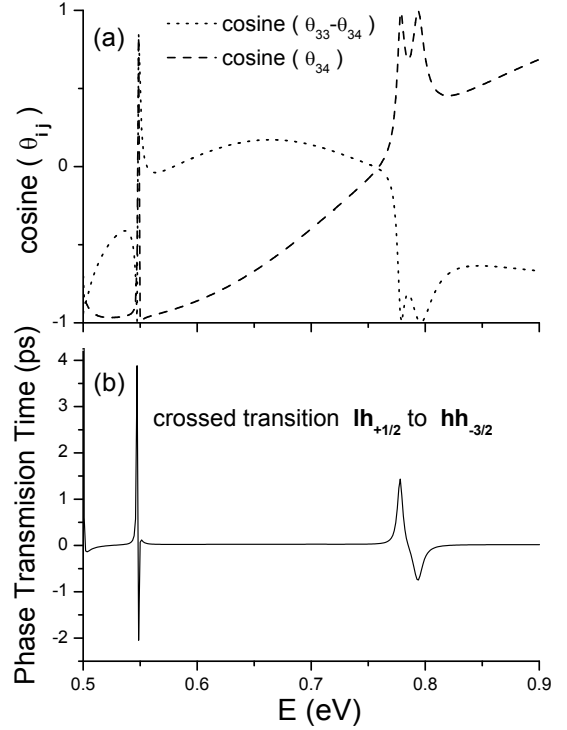}
  \caption{\label{FigCros3} The bottom panel plots tunneling phase time for the
 crossed-transition $j:3 \rightarrow 4:i$ through a single barrier heterostructure at
 $\kappa_{\textsc t}= 0.001$ \AA$^{-1}$, with $L_{b}=20$ \AA $\,$ and
 $V_{b}=0.498$ eV. Top panel shows the cosine of phase shift for the same
 transition $j:3 \rightarrow 4:i$ (dashed line) where the first number stands for the input channel $j$ and the other stands for the output one $i$. In dotted line the same magnitude is shown for the direct path $j:3 \rightarrow 3:i$ with respect to the crossed path $j:3 \rightarrow 4:i$.}
\end{figure}

The porpoise of Figure \ref{FigCros3}, is to briefly analyze further anomalous features in the transmission phase time spectra. Fig.\ref{FigCros3} displays in panel (a) the $cos(\theta_{ij})_{n=1}$ for the crossed
$lh-hh$ transition $lh_{+ 1/2}\rightarrow hh_{- 3/2}$ (dashed line, $i:3 \rightarrow
4:j$). In panel (a) with dotted line, we plot the difference of the the same magnitude for
the path $i:3 \rightarrow 3:j$ with respect to the path $i:3 \rightarrow 4:j$.
The bottom panel (b) plots tunneling phase time for the crossed-transition $i:3
\rightarrow 4:j$ through a single barrier heterostructure. In the presence of
interference effects ($\kappa_{\textsc t} \neq 0$) at $E \approx 0.55$ eV, the phase
changes abruptly when $\Delta\theta \approx \mb{\large $\pi$}$ radians, as can be seen
from the dashed line in the top panel of Figure \ref{FigCros3}. Both curve in panel
(a) show a clear correspondence of abrupt changes in $\cos \theta_{ij}$ with observed
peaks(dips) of $\mb{\large $\tau$}_{ij}$, which seem to be an anomalous behavior of the
phase time, as they resemble Fano-like resonances due theirs asymmetry hallmark \cite{Fano61}. In the neighborhood of $0.8$ eV [see panel (b)] we have found Fano-like
positive (negative) peaks (dips) of $\mb{\large $\tau$}_{ij}$ with $i \neq j$, displayed
in the bottom panel of Figure \ref{FigCros3}. They are associated with changes of the phase in almost $\mb{\large $\pi/2$} ($\mb{\large $\pi/3$}$)$ radians, and are plotted in the top
panel with dashed line. We also have observed, although will not be shown here, local
increments of transmission coefficients in the energy vicinity, where negative dips
appeared, though regretfully they need more investigation to derive proper
conclusions.

Most interestingly, though, we found indications that might be connected to possible giant conductance events in the single-cell case, which we omit graphically here because of brevity. There were found two prompt shifts in transmission phase amplitude, equivalent to an even multiple of $\mb{\large$\pi$}$ ($|\Delta\theta_{ij}|=6\mb{\large $\pi$}$) at $\kappa_{\textsc t} \approx 0.01$ \AA$^{-1}$, similar to those plotted in Figure \ref{FigGG}. One of them, appear in $lh_{+1/2}\rightarrow lh_{-1/2}$ intra-band transition ($i:3 \rightarrow 2:j$) when $L_{b}$ varies within $[10-20]$ \AA$\,$ range. The second one, was observed through $hh_{-3/2}\rightarrow lh_{+1/2}$ inter-band transition ($i:1 \rightarrow 4:j$) when $L_{b}$ varies within $[120-130]$ \AA$\,$ range. Then, maxima of $G_{j}$ could be expected for those crossed paths, not only varying the incident energy, but the barrier thickness as well. However, a clear resonant mechanism in this case remains a puzzle and further investigations need to be addressed to clarify wether the effect of giant conductance for holes, is expected to arise under valence-band coupling phenomena during transport trough a single-cell system.

\section{Concluding Remarks}
\label{conclusions}

Relevant and striking quantum transport features of holes, directly connected to the
phase, the phase shift and the phase resonant interference phenomena are: the giant
conductance-resistance events, the Hartman effect, and the earlier phase propagation in
the barrier. Several results nicely resemble those previously reported for electrons
\cite{PRLsuperlatticetimes}, while some predictions of the MSA model are in acceptable
qualitative compliance with the experiments \cite{EWBL85,Drag03,Ten96} and with
theoretical predictions \cite{Landauer89,PPP02}. Theoretical evidences of the giant
conductance phenomena for decoupled hole transmission through DBRT were exhibited. In this
case a trustworthy resonant mechanism supporting this anomaly, is the alignment of the
incident propagation energy with one of the quasi-bond hole states of the embedded quantum
well in the DBRT. This mechanism was confirmed with an independent numerical calculations of several quasi-bound levels. The giant conductance effect vanishes in the presence of valence-band particles coupling and by tuning the number of superlattice layers as well. The hole phase transmission time for a $(GaAs/Al_{0.3}Ga_{0.7}As/GaAs)^{n}$ superlattice, shows a resonant mini-bands structure in the scattering regions, having the free motion as an inferior limit time. Moreover, in the forbidden regions the holes propagation became away faster and was shown $\tau_{ii}$ to be inferiorly bounded by the single-cell phase time $\tau_{1}$. In the DBRT mini-gaps regions, $\tau_{ii}$ moves forward from the free motion time, suggesting a more speedily phase propagation within the barrier. The Hartman's classic
prediction is observed robust for null and finite hole band mixing.
Was presented an alternate-selective confinement strength independently for both flavors of holes, implying filter-like effects on the effective mass \textit{via} the manipulation of the incident energy in the uncoupled hole regime.

\section*{Acknowledgments}
We thanks Dr. H. Rodríguez-Coppola for clue suggestions in some numerical treatments. One of the authors (L.D-C) gratefully acknowledges the Visiting Academic Program of the UIA/M\'{e}xico and the facilities of the IFSC-USP/S\~{a}o Carlos, Brasil.


\begin{thebibliography}{00}
\bibitem{Com1} We have obtained the amplitude of the exponential fitting
function, for the phase transmission time with different sections of barrier
thickness, to be constant starting from $55$ \AA.
\end{thebibliography}
\end{document}